\documentclass[twocolumn,a4paper,showpacs,preprintnumbers,aip,jap,amssymb,amsmath,showkeys,10pt]{revtex4-1}

\usepackage[francais,english]{babel}
\usepackage[utf8]{inputenc}
\usepackage[T1]{fontenc}
\usepackage{fouriernc}

\usepackage{amsmath}
\usepackage{textcomp}
\usepackage{natbib}
\usepackage{graphicx}
\usepackage{dcolumn}
\usepackage{bm}

\begin{document}

\title[18pt]{Broadband ferromagnetic resonance characterization of GaMnAs thin films}

\author{A. Ben Hamida}
\affiliation{\rm{Physikalisch-Technische Bundesanstalt}, \emph{Bundesallee 100, D-38116 Braunschweig, Germany}}
\author{S. Sievers}
\affiliation{\rm{Physikalisch-Technische Bundesanstalt}, \emph{Bundesallee 100, D-38116 Braunschweig, Germany}}
\author{K. Pierz}
\affiliation{\rm{Physikalisch-Technische Bundesanstalt}, \emph{Bundesallee 100, D-38116 Braunschweig, Germany}}
\author{H.W. Schumacher}
\affiliation{\rm{Physikalisch-Technische Bundesanstalt}, \emph{Bundesallee 100, D-38116 Braunschweig, Germany}}

\begin{abstract}

The precessional magnetization dynamics of GaMnAs thin films are characterized by broadband network analyzer ferromagnetic resonance (FMR) in a coplanar geometry at cryogenic temperatures. The FMR frequencies are characterized as function of in-plane field angle and field amplitude. Using an extended Kittel model of the FMR dispersion the magnetic film parameters such as saturation magnetization and anisotropies are derived. The modification of the FMR behavior and of the magnetic parameters of the thin film upon annealing is analyzed.
\end{abstract}

\maketitle

\section{Introduction}

Precessional magnetization dynamics of magnetic thin films and nanostructures are highly relevant for magnetic device applications. For example the minimum magnetization reversal times and hence the ultimate data rate of a magnetic memory devices is directly determined by the precession frequency \cite{Kaka_APL80_2958_2002, Gerrits_Nature418_509_2002, Schumacher_PRL90_017201_2003, Schumacher_PRL90_017204_2003}. Ferromagnetic semiconductors are a particularly promising class of magnetic materials as they could offer the combination of magnetic memory and semiconductor logic functions in the same material. Presently, GaMnAs can be considered the most prominent prototype of diluted ferromagnetic semiconductors with well determined material parameters \cite{Jungwirth_RMP78_809_2006}. The precessional magnetization dynamics of GaMnAs thin films and devices have been characterized by different techniques. Time resolved precessional dynamics have been studied by all-optical pump-probe mageto optics using fs lasers \cite{Qi_APL91_112506_2007, Rozkotova_APL92_122507_2008, Rozkotova_APL93_232505_2008, Hashimoto_PRL100_067202_2008, Zhu_APL94_142109_2009} or by time resolved magneto-optical characterization upon electrical excitation \cite{Hoffmann_PRB80_054417_2009, Woltersdorf_PRB87_054422_2013}. Furthermore low-temperature cavity ferromagnetic resonance (FMR) has been used to investigate anisotropies and linewidths \cite{Sasaki_JAP91_7484_2002, Liu_PRB67_205204_2003} as well as spin wave resonances \cite{Goennenwein_APL82_730_2003}. In addition electrical measurements based on photovoltage detection \cite{Wirthmann_APL92_232106_2008} or on spin-orbit ferromagnetic resonance \cite{Fang_NatureNano6_413_2011} have been tested.\\
 
Over the last years broadband vector network analyzer based ferromagnetic resonance (VNA-FMR) \cite{Sangita_JAP99_093909_2006} using coplanar wave guides as inductive antennas has become a versatile tool for the simple and fast all electrical characterization of precessional dynamics of various magnetic thin films and multilayers \cite{Santiago_JAP110_023906_2011, Santiago_JAP109_013907_2011}. In principle low-temperature VNA-FMR \cite{Sierra_APL93_172510_2008} should also be suitable for the electrical characterization of ferromagnetic semiconductors such as GaMnAs. However up to now VNA-FMR based measurements of the precessional magnetization dynamics of GaMnAs have proven difficult to achieve. This is due to the low saturation magnetization of GaMnAs in combination with strong crystalline anisotropies. Both properties lead to a rather weak inductive signal which could be easily masked by the noise of the high bandwidth measurement electronics.\\

Here we present broadband coplanar VNA-FMR measurements of the precessional dynamics of GaMnAs thin films in a cryogenic environment. For sufficiently high excitation powers a clear precessional signal is observed in the VNA-FMR spectra. The field and angular dependence of the FMR peaks can be well described by a Kittel model taking into account the different anisotropy components of \mbox{GaMnAs}. The precessional dynamics and material parameters of an as-grown and an annealed thin film are analyzed and compared. 

\section{Experimental setup and procedure}

GaMnAs layers of 100$\ $nm thickness were grown in a low-temperature MBE environment on a 2 inch semi-insulating GaAs(001) wafer at temperatures of $T_g$ = 240°C and 220°C, respectively. Details of the growth and annealing procedure can be found elsewhere \cite{BenHamida_JMSJ36_49_2012}. Samples of 10$\ $mm x 10$\ $mm were cut from the wafers as well as 5$\ $mm x 5$\ $mm pieces for superconducting quantum interference device (SQUID) magnetometry. For the as-grown sample of $T_g$ = 240°C a saturation magnetization $M_S$ = 30$\ $mT is measured. The sample with $T_g$ = 220°C was annealed at 200°C for 18$\ $h in ambient air  resulting in an increased saturation magnetization of $M_S$ = 74$\ $mT.\\

The setup for inductive FMR characterization of the samples is described with respect to Fig. 1. The experiments are carried out in a variable temperature insert of a commercial He cryostat allowing to vary the sample temperature from $T_S$ = 1.5 … 250 K. All FMR measurements of this work were carried out at fixed sample temperature of $T_S$ = 10 K. The cryostat is equipped with a three axial superconducting vector magnet allowing application of static magnetic vector fields $\mu_0H_s$ = $\mu_0(H_x,H_y,H_z)$ up to 1 T amplitude and arbitrary orientation. For FMR measurements the 5x5 mm$^2$ samples are placed on the center of a coplanar waveguide (CPW). Details of the design and high frequency properties of the CPW can be found elsewhere\cite{Sievers_IEEETransMag49_58_2013}. Both ends of the CPW are connected to the two ports of a 24 GHz bandwidth VNA via 18 GHz bandwidth coaxial lines of about 1.5 m length. Note that the rather long length of the coaxial lines is determined by the dimensions of the cryostat. Inside the cryostat the CPW substrate is oriented in the $xy$-plane with the CPW line running along $x$ and high frequency (HF) field generation along $y$ ($H_{HF}$ || $H_y$). In the present experiments in-plane magnetic vector fields $\mu_0(H_x,H_y)$ up to 0.5 T amplitude are applied in the sample plane ($H_z$ = 0). As sketched in Fig. 1(a) the GaMnAs samples are placed diagonally on the CPW with the [010] crystalline axis oriented parallel to $H_{HF}$ and the [100] axis along $x$.\\

For an FMR measurement at a given static vector field $(H_x,H_y)$, the frequency output of the VNA is swept between 1 and 18 GHz and the forward scattering signal $S_{21}$ is measured. To maximize the weak inductive FMR signal the maximum output power of 20 dBm is applied. The HF output signal generates a HF excitation field $H_{HF}$ around the CPW center conductor line and thus in the GaMnAs thin film sample. Under resonance conditions the GaMnAs magnetization is excited into FMR precession and the signal transmission is reduced leading to a Lorentzian absortion line in the VNA sweep.  To enhance the visibility of the FMR signal a VNR reference measurement $S_{21,ref}$ is carried out at a non-resonant field. The normalized transmission signal T is then deduced by subtracting the reference VNA sweep from the actual measurement data. From the resonance peak the FMR frequency $f_{FMR}$ and the absorption line width $\Delta{f_{FMR}}$ are derived. Note that for certain applied static fields only weak resonance peaks were found making a reliable linewidth analysis impossible. By variation of the applied static field the FMR properties were measured as function of the field vector $(H_x,H_y)$.\\

\begin{figure}
\centering\includegraphics[width=\linewidth]{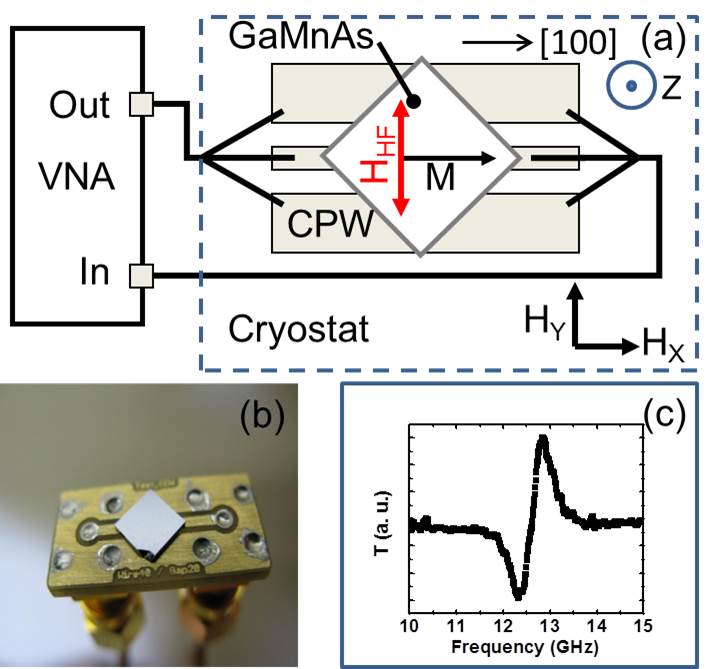}
\renewcommand{\figurename}{Fig.}
\caption{(a) Experimental Setup: Sketch of the sample position on the coplanar waveguide and connexions with the VNA. (b) Photograph of the sample on top of the coplanar waveguide. (c) Experimental FMR curve at $H$ = 0.32$\ $T. The absorption line has the shape of an asymmetric Lorentzian around $f_{FMR}$ = 12.6 GHz with linewidth $\Delta{f}$ = 0.5 GHz.}
\end{figure}

\section{Data Analysis}

A detailed review on FMR in GaMnAs has been given by Liu and Furdyna \cite{Liu_JPCM18_R245_2006}. In an FMR experiment, the magnetization $\vec{M}=M_s(\sin\theta\cos\phi,\sin\theta\sin\phi,cos\theta)$ (see Fig. 2 for details of the angle nomenclature) of the film precesses around its equilibrium position with the FMR frequency $f_{FMR}$. Sweeping the value of the applied microwave frequency $f_{HF}$ at a fixed magnetic field $\vec{H}=H(\sin\theta_H\cos\phi_H,\sin\theta_H\sin\phi_H,cos\theta_H)$ (cp. Fig. 2), the resonance condition will be satisfied at $f_{FMR} = f_{HF}$. The condition is given by: 
\begin{equation}
(\frac{2\pi{f_{HF}}}{\gamma})^2 = \frac{1}{(2\pi{M_s}\sin\theta)^2}[\frac{\partial^2{F}}{\partial\theta^2}\frac{\partial^2{F}}{\partial\phi^2}-(\frac{\partial^2{F}}{\partial\theta\partial\phi})^2]
\end{equation}
where $\gamma = 1.76*10^{11}$ denotes the gyromagnetic ratio and $F$ is the free magnetic energy. The expression of $F$ for a thin film with crystalline and uniaxial anisotropy is given by:
\begin{eqnarray}
F &=& -\mu\vec{M}\vec{H} + \frac{\mu}{2}M_sM_{eff}\cos^2\theta - K_{u1}\sin^2\theta\cos^2(\phi-\Omega_1) \nonumber \\
  &+& \frac{K_{c1}}{4}(\sin^2(2\theta) + \sin^4\theta\sin^2(2\phi-2\Omega_c))
\end{eqnarray} 
where $\mu = 4\pi{10^{-7}}$ (SI units). $M_{eff}$ is the effective magnetization of the thin film. $K_{u1}$ is the in-plane uniaxial anisotropy constant corresponding to the easy axis orientation $\Omega_1$ with respect to the crytallographic direction [100]. $K_{c1}$ is the in-plane cubic anisotropy constant. $\Omega_c$ is the in-plane orientation of the cubic easy axis with respect to [100].\\

\begin{figure}[ht]
\centering\includegraphics[scale=0.5]{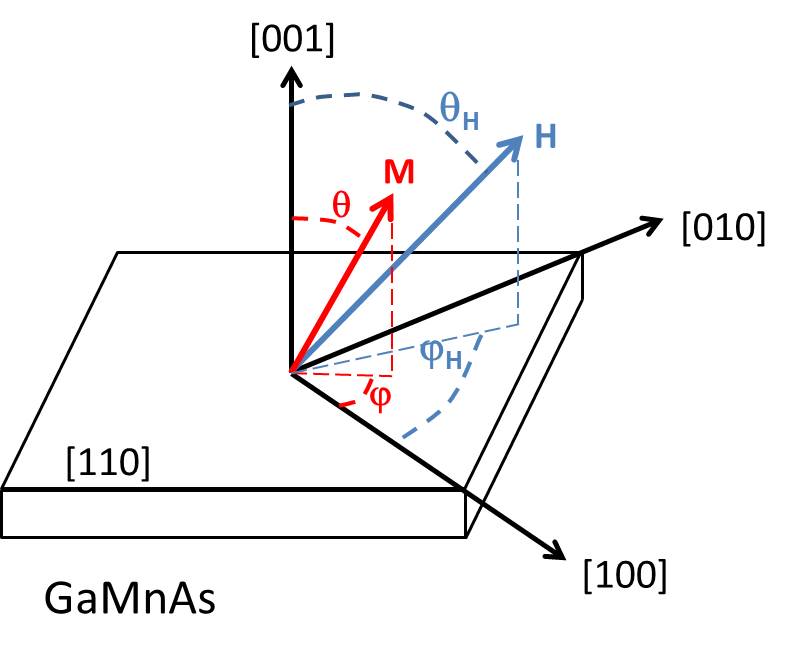}
\renewcommand{\figurename}{Fig.}
\caption{GaMnAs sample: Configuration of the magnetization $\vec{M}$ and the magnetic field with respect to the crytallographic directions. $\phi$ and $\phi_H$ are the in-plane angles of $\vec{M}$ and $\vec{H}$, respectively, as measured from the [100] orientation. $\theta$ and $\theta_H$ are the normal angles of $\vec{M}$ and $\vec{H}$, respectively.}
\end{figure}

\begin{table*}[ht]
\begin{center}
\begin{tabular}{|p{0.1\linewidth}|p{0.08\linewidth}|p{0.08\linewidth}|p{0.08\linewidth}|p{0.08\linewidth}|p{0.08\linewidth}|p{0.08\linewidth}|p{0.05\linewidth}|p{0.05\linewidth}|p{0.05\linewidth}|}
\hline
Sample & $M_S$ & $M_{eff}$ & $H_{u1}$ & $\Omega_1$ & $H_{c1}$ & $\Omega_c$ \\ \hline
annealed & 74 mT & 30 mT & 70 mT & 45° & 85 mT & 0°  \\ \hline
as-grwon & 30 mT & 130 mT & -20 mT& 20° & 100 mT & -20° \\ \hline
\end{tabular}
\end{center}
\caption{Experimental derived values of the magnetic parameters for annealed and as-grown 100 nm thick GaMnAs samples.}
\end{table*}

During our experiments, the external field $\vec{H}$ is applied in-plane and the film magnetization $\vec{M}$ is assumed to stay also in-plane ($\theta_H = \theta = 90$°). At a given direction of $\vec{H}$, the resonance is then obtained by numerically solving the above equation at the equilibrium position of $\vec{M}$, for $\frac{\partial{F}}{\partial\phi}=0$. The anisotropy parameters of the GaMnAs thin films can then be derived for example from a fit to the measured angular dependence of the precession frequency at fixed field amplitude H. Fig 3 shows $f_{FMR}$ as function of the in-plane field angle $\phi_H$ for applied field amplitude of $\mu_0H$ = 0.2 T (green) and 0.3 T (red). The symbols represent the measured data whereas the lines are the fits according to the above FMR model. Two sets of data taken on the two different samples are shown. The upper panel (a) shows the data of the annealed sample whereas in (b) the data of the as-grown sample are shown.\\

\begin{figure}[ht]
\centering\includegraphics[width=\linewidth]{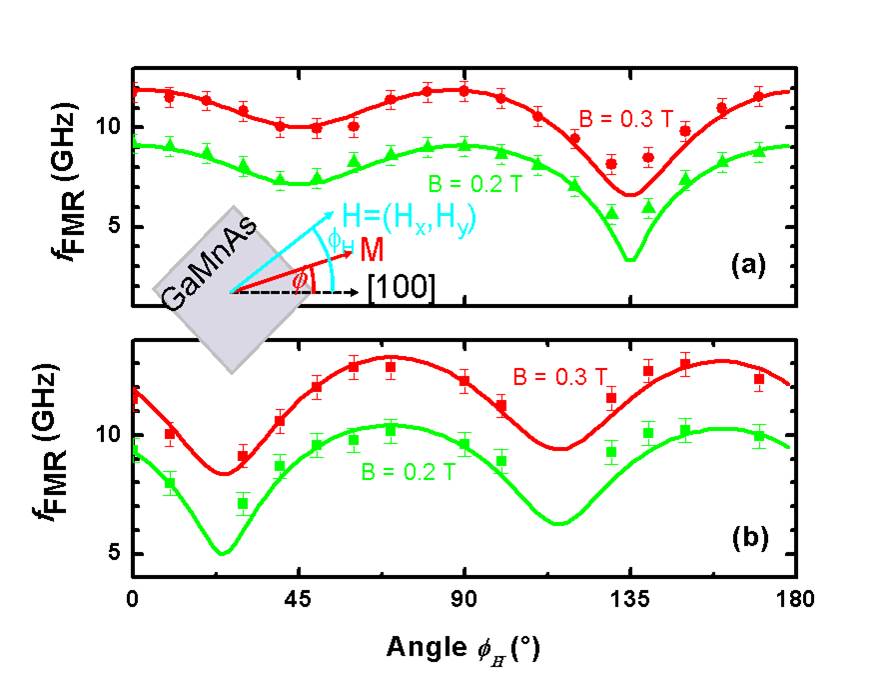}
\renewcommand{\figurename}{Fig.}
\caption{Angular dependence of the precession frequency for two field amplitudes of 0.2 T (green) and 0.3 T (red). The symbols are experimental data. The solid lines are the fit. (a) Annealed sample, (b) as-grown sample.}
\end{figure}

The data of both samples can be well described by a fit to the above model taking into account a thin film with crystalline in-plane cubic and uniaxial anisotropy. The values of the derived anisotropies for the best fit to the data are regrouped in Table 1. Note that in Table 1 the anisotropy fields $H_i = 2K_i/M_S$ are given instead of the anisotropy constants $K_i$.\\

Note that the values of the saturation magnetization $M_S$ are based on the SQUID measurements and are not derived from fitting. Figure 3 shows that the model based on the above parameters well describe the experimental data both of the annealed and the as-grown sample. The derived values of the magnetic parameters are very reasonable when compared to literature values of thin film anisotropies derived by conventional cavity FMR experiments of GaMnAs thin films \cite{Liu_JPCM18_R245_2006}. The comparison of the two parameter sets shows the strong impack of annealing on all magnetic parameters from saturation magnetization to the various anisotropy terms.\\  

The broadband coplanar FMR setup also allows to derive experimental values of $f_{FMR}$ as function of field amplitude $H$ for a fixed in-plane angle $\phi_H$. Such field dependent data is shown in Fig. 4 for selected field angles for the (a) annealed and (b) as-grown samples. The symbols again represent the data whereas the lines are the model fit. The measured data is again well described by the model fit confirming the feasibility of the derived parameters.\\

\begin{figure}[ht]
\centering\includegraphics[width=\linewidth]{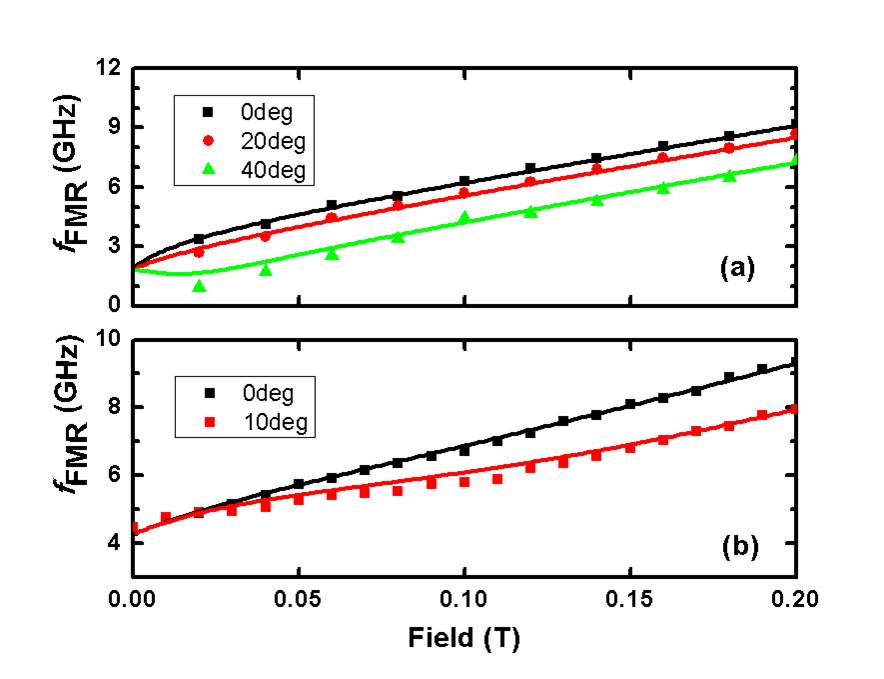}
\renewcommand{\figurename}{Fig.}
\caption{Field dependence of the precession frequency for different field orientations $\phi_H$. The points are experimental data. The solid lines are the fit. (a) Annealed sample, (b) as-grown sample.}
\end{figure}

As mentioned above the linewidth could not be systematically analyzed from the VNA-FMR data. However for selected data points a sufficiently clear resonance peak allowed a reliable linewidth analysis. From this linewidth data a Gilbert damping parameter of $\alpha = 0.018$ was derived for the annealed sample. This value is in good agreement with the literature values of annealed samples derived by X-Band ferromagnetic spectroscopy \cite{Khazen_PRB77_165204_2008}. It is however worth noting that this value is quite different from literature data derived from time resolved optical pump probe experiments \cite{Qi_APL91_112506_2007} for an as-grown sample (the values of $\alpha$ ranged from 0.12 to 0.21). The reason of this strong deviation of the literature values of the damping derived by different methods can presently only be subject of speculation. However, it might be related to different applied fields and hence to different contributions of extrinsic line broadening in the experiments \cite{Counil_JAP95_5646_2004}. Furthermore the difference could be related to inhomogeneous sample properties \cite{Woltersdorf_PRB87_054422_2013}.

\section{Conclusion}

Concluding we have demonstrated the suitability of broadband network analyze based FMR for the characterization of the precessional dynamics of GaMnAs thin films. A coplanar inductive antenna was used to excite and detect the precessional signal of 100 nm thick annealed and as-grown GaMnAs thin films with saturation magnetization down to 30 mT. The field and angular dependence of the FMR frequency could be well described by a model taking into accound the different thin film anisotropy terms. The simple and yet powerful setup could in the future allow investigations of more complex systems such as of the coupled dynamics of GaMnAs based tunnel junctions and multilayers.

\section{Acknowledgments}

The work was supported by DFG SPP \mbox{Semiconductor} Spintronics and EMRP JRP IND08 MetMags and JRP EXL04 SpinCal. The EMRP is jointly funded by the EMRP participating countries within EURAMET and the EU.
 
\renewcommand\refname{\lowercase{\uppercase{R}eferences}} 

\end{document}